# Well-Ordered In Adatoms at the In$_2$O$_3$(111) Surface Created by Fe Deposition

*Margareta Wagner[1,*], Peter Lackner[1], Steffen Seiler[2], Stefan Gerhold[1], Jacek Osiecki[3], Karina Schulte[3], Lynn A. Boatner[4], Michael Schmid[1], Bernd Meyer[2], and Ulrike Diebold[1]*

[1] Institute of Applied Physics, Vienna University of Technology, Wiedner Hauptstraße 8-10/134, 1040 Vienna, Austria

[2] Interdisciplinary Center for Molecular Materials and Computer-Chemistry-Center, Friedrich-Alexander-University Erlangen-Nürnberg, Nägelsbachstraße 25, 91052 Erlangen, Germany

[3] MAX IV Laboratory, Lund University, Ole Römers väg 1, 223 63 Lund, Sweden

[4] Materials Science and Technology Division, Oak Ridge National Laboratory, Oak Ridge, Tennessee 37831, USA

*corresponding author: wagner@iap.tuwien.ac.at

*Abstract*

Metal deposition on oxide surfaces usually results in adatoms, clusters, or islands of the deposited material, where defects in the surface often act as nucleation centers. Here an alternate configuration is reported. After the vapor deposition of Fe on the In$_2$O$_3$(111) surface at room temperature, ordered adatoms are observed with scanning tunneling microscopy. These are identical to the In adatoms that form when the sample is reduced by heating in ultrahigh vacuum. Density functional theory (DFT) calculations confirm that Fe interchanges with In in the topmost layer, pushing the excess In atoms to the surface where they arrange as a well-ordered adatom array.

Metal-oxide surfaces are often functionalized by vapor-depositing small amounts of a foreign metal. For relatively noble metals (e.g., the coinage elements Ag, Au, and Pt, Pd) deposition at ambient temperatures usually results in aggregation into small nanoclusters that may nucleate at special sites such as surface defects. While some charge transfer between the overlayer and the metal oxide does occur for these elements (and can lead to interesting effects [1, 2]), the substrate itself is generally not affected. This situation changes when the deposited metal has a high affinity for oxygen; it can react with the surface, which leads to a re-arrangement of the local atomic structure. While the situation is well investigated for the relatively unreactive late-transition metals on many different oxide supports (which are often used as models in heterogeneous catalysis), surprisingly little is known about what happens at the atomic scale when reactive metals are deposited.

Two recent Scanning Tunneling Microscopy (STM) studies have shown that the foreign metal can be incorporated into the near-surface region [3, 4]. For ultrathin NaCl films on Au(111), it was found that Co atoms replace either Na or Cl; the excess substrate atoms are assumed to diffuse to the Au substrate (Na) or the borders of the NaCl islands (both Na and Cl). When reactive metals (Ti, Co, Cr, Zr) were deposited on $Fe_3O_4$(001) [4], they filled up the Fe vacancies in the subsurface layer, which are an inherent part of this materials' surface structure [5]. Macroscopic measurements of ultrathin films show a spreading of the reactive overlayer metal (rather than clustering), and a loss of long-range geometric structure – also indicating a re-arrangement of the substrate [6]. This has been found, e.g., on $TiO_2$(110) [7] for the growth of Al and many transition metals of the groups IVB-VIIB including Ti, Hf, Cr, V, Mn and Fe, and on Fe(001)-p(1×1)O for Cr, V, Nb, Mo but also Ba and Ni [8].

Deposition of electropositive elements also results in a re-arrangement of charge: the excess electrons that are brought into the system reduce the substrate cation. Such an electronic reduction of the system can also lead to structural changes. E.g., it is well-established that the much-investigated material rutile $TiO_2$(110) forms a (1×2) reconstruction upon reduction that is brought upon by the preferential sputtering of oxygen or heating in vacuum [9]. When a system is doped with an aliovalent atom, incorporation into the lattice is often accompanied by the formation of compensating defects such as O vacancies [10, 11]. In the case of Fe reactively deposited on rutile $TiO_2$(011)(2×1), a reconstruction into an ordered mixed oxide surface layer with Fe(II)Ti(IV)$O_3$ stoichiometry is observed [12].

This work investigates single-crystal indium oxide ($In_2O_3$). This transparent conductive oxide (TCO) is often doped with tin, and is then commonly referred to as indium tin oxide (ITO). It is widely used as hole-injecting contact material in opto-electronic devices, and also finds application as gas sensor, e.g., for the detection of CO or ethanol [13, 14]. Compared to other technologically important oxides, little is known about its surfaces – although recent investigations have led to interesting results, e.g., the determination of the fundamental band gap as 2.93 eV [15, 16].

Here we focus our attention on the (111) surface of $In_2O_3$, which is both nonpolar and the thermodynamically most stable facet. It exhibits a simple 'bulk-terminated' (1 × 1) configuration [17]. When reduced by heating in vacuum, $In_2O_3$(111) responds in a rather unusual way: instead of oxygen vacancies, excess In atoms appear on the surface. Interestingly, this In is present in the form of isolated adatoms that sit at specific sites in the surface unit cell (see Fig. 1). This has been shown recently, where a mild reduction [annealing in ultrahigh vacuum (UHV) to 500°C] resulted in a full layer of well-ordered (native) In adatoms on $In_2O_3$(111) [18]. A step-wise oxidation of the reduced surface decreased the number of isolated adatoms, which transformed into single-layer $In_2O_3$(111) islands on terraces. The preference for forming In adatoms rather than oxygen vacancies was also confirmed by DFT calculations [18].

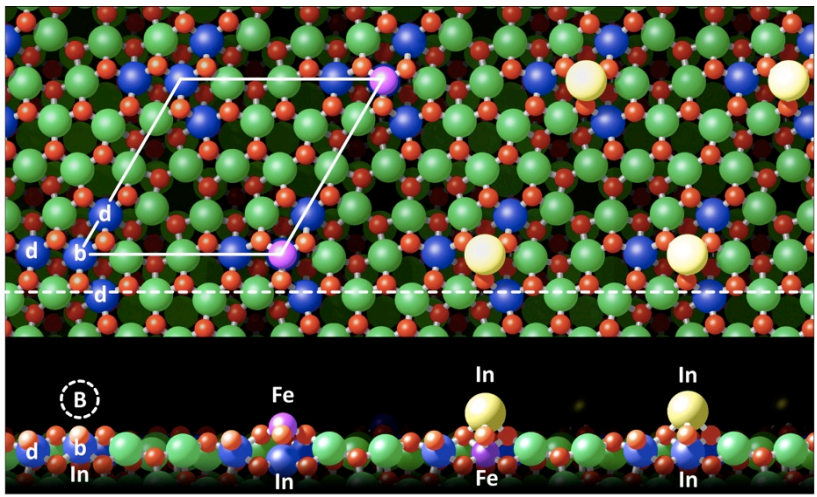

Fig. 1: Surface structure of $In_2O_3$(111), top and side view, with the unit cell indicated. The oxidized surface is essentially bulk-terminated (extreme left), whereas the reduced $In_2O_3$ surface contains In adatoms at site 'B' (extreme right). When Fe is deposited at room temperature, it does not stay at the most-favorable adsorption site 'B' (left), but replaces an In atom in the lattice (preferentially 'b') and "pushes" the In atom to the surface, where it adsorbs as an adatom in 'B' (right). Oxygen atoms are red, five-fold (5c) and six-fold (6c) coordinated Indium atoms are green and blue, respectively; In adatoms are yellow and Fe is pink.

In this work we report the initial stage of Fe grown on $In_2O_3$(111) at room temperature. Measurements with STM, X-ray photoelectron spectroscopy (XPS), low-energy ion scattering (LEIS) and Auger electron spectroscopy (AES) are complemented by density functional theory (DFT) calculations. We found that Fe atoms are incorporated into the surface by replacing In atoms. These excess In atoms are "pushed" to the surface where they prefer a specific position within the surface unit cell, thus resulting in an ordered array of single In adatoms. Their presence will modify materials properties such as the electronic structure, gap states, work function, or chemical reactivity. This adsorption mechanism has not been reported previously on any metal/oxide system, it is well possible that other compounds may react in a similar way.

The STM measurements were carried out in a UHV system (base pressure $<3\times10^{-10}$ mbar) described elsewhere [18], equipped with an electron beam evaporator (Focus) for Fe deposition. A different UHV system, described in Ref. [19], was used for AES and LEIS studies. The XPS und UPS measurements were performed at the beam line I311 of the Max IV laboratory in Lund, Sweden. The $In_2O_3$ single crystals and the sample preparation procedure are described in Refs. [18, 20]. The lattice parameter of $In_2O_3$(111) is 1.430 nm; a single-layer step height measures 292 pm. The surface unit cell is an O-In-O tri-layer containing 16 $In^{3+}$ ions (4 and 12 with 6- and 5-fold coordination, respectively) and 24 $O^{2-}$ ions, half in 4-fold coordination below and half in 3-fold coordination above the In layer (see Fig. 1). Full periodicity perpendicular to the surface requires six trilayers. At the surface, the In(6c)/O(3c) (c = coordinated) regions appear as dark triangles in STM images [17, 18]. Reducing the surface by annealing in UHV (500°C) stabilizes In adatoms (see Fig. 1) in the 3-fold hollow sites labeled 'B' [21] above the central In(6c) atom 'b', with a coverage of one adatom per unit cell, see Ref. [18]. Iron was evaporated at room temperature (RT) onto the oxidized (no In adatoms) or the partially reduced (covered with some native In adatoms) $In_2O_3$(111) surface. Note that one monolayer (ML) refers to one adatom per surface unit cell, i.e., a density of $5.65\times10^{13}$ $cm^{-2}$.

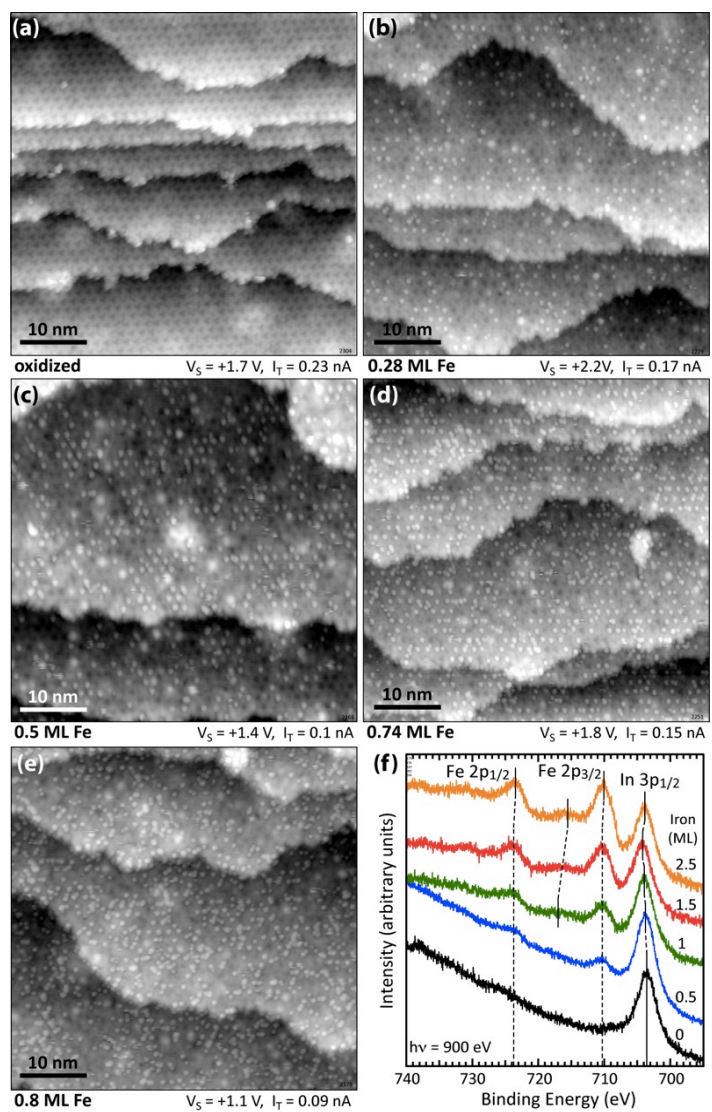

Fig. 2: The oxidized, bulk-terminated $In_2O_3$(111) surface after Fe deposition at room temperature. STM images of (a) the clean surface, and with (b-e) 0.28, 0.5, 0.74 and 0.8 ML Fe deposited at room temperature. (f) XPS spectra of the Fe 2p core levels starting with the clean surface (black curve at the bottom) up to 2.5 ML of Fe. The Fe 2p levels are marked by dotted lines to guide the eye. The In 3p peak shows small shifts due to band bending.

After iron deposition, STM data shows adatoms on the $In_2O_3$(111) surface, see Fig. 2. They occupy the same adsorption site B as recently observed for native In adatoms after reduction [18]. Figs. 2(a-e) display the oxidized surface and after the deposition of 0.28, 0.5, 0.74 and 0.8 ML of Fe, respectively. Up to a coverage of 0.5 ML the adatoms are ordered, and the dark triangles of "unoccupied" unit cells are easily distinguished. Increasing the coverage leads to disordered adatoms, and eventually to large clusters and 3-dimensional growth (see Supplemental Material [22]). The photoemission spectra in Fig. 2(f) show the Fe

2p core levels with increasing amounts of iron. The spectra were acquired in grazing emission (55° from the surface normal). The Fe $2p_{3/2}$ ($2p_{1/2}$) peak of 0.5 ML Fe (ordered regime) is located at ~710.6 eV (~723.8 eV), i.e., shifted by +4.3 eV with respect to metallic iron (706.3 eV [31]). The satellite at ~717 eV, associated with $Fe^{3+}$, starts to become visible at 1 ML. At higher coverages (disordered cluster regime), the satellite shifts to lower binding energy and a shoulder on the low binding energy side of Fe $2p_{3/2}$ is discernable – indicating the presence of Fe in various oxidation states. In the valence-band photoemission a gap state around 2 eV is observed (see Supplemental Material [22]). The shifts of the In $3p_{1/2}$ peak to slightly higher binding energies, also present in the valence band structure (see Supplemental Material [22]), is assigned to band bending.

From the STM and XPS results one could conclude that Fe adsorbs as adatoms, which are ordered at very low coverages and disorder when they fail to find an unoccupied unit cell within their diffusion length. The presence of 0.5 ML Fe in the top few layers of the sample was also confirmed with AES and LEIS measurements. In LEIS spectra the Fe signal vanished quickly due to mild sputtering effects (see Supplemental Material [22]).

It is puzzling, however, that the STM images and corrugation profiles of the reduced $In_2O_3$(111) surface with native In adatoms look identical to the oxidized $In_2O_3$(111) surface after Fe deposition, see Fig. 3. Here, a partially reduced surface was prepared by annealing at 500 °C in $1.5\times10^{-8}$ mbar $O_2$ for 10 min, and by keeping the sample in oxygen during cooling down to 250 °C. This procedure results in a coverage of native In adatoms of 0.35 ML, see Fig. 3(a). Their apparent height is ~150 pm – in agreement with Ref. [18]. After the deposition of 0.27 ML Fe at RT (see Fig. 3(b)), the total coverage of adatoms, i.e., the sum of native In adatoms and Fe induced adatoms, is 0.62 ML. The linescan in Fig. 3(c) shows the heights of the adatoms. Bias-voltage-dependent STM images with sample voltages ranging from +2.2 V to -1.8 V are given in the Supplemental Material [22]. Although the apparent height changes slightly with the tip condition and bias, only one type of adatoms is consistently observed. This leads to the conclusion that the Fe-induced adatoms (of the ordered regime) could be In. This would imply that Fe could replace In in the lattice, with the surplus In pushed out and becoming adatoms.

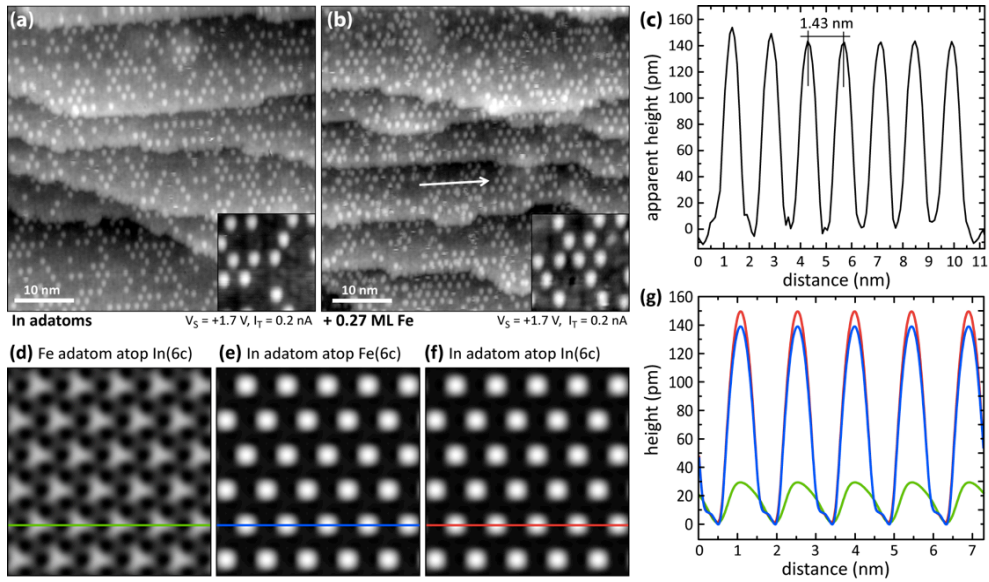

Fig. 3: Comparison of native In adatoms with Fe-induced adatoms. (a) Partially reduced In$_2$O$_3$(111) surface with native In adatoms only, and (b) the same surface after deposition of 0.27 ML Fe at room temperature, i.e., with native In adatoms and Fe induced adatoms. (c) Linescan along the arrow indicated in panel (b). STM calculations (Bardeen approach) of the most stable configuration for (d) Fe adatom adsorption and (e) Fe substituting the central In(6c) atom (labeled 'b' in Fig. 1), compared to (f) native In adatoms. (g) Simulated STM height profiles across the adatoms of the structures shown in (d-f).

To test this hypothesis, we performed DFT calculations and considered various configurations for an Fe atom interacting with the In$_2$O$_3$(111) surface – including adsorption on the surface, incorporation as interstitials, and replacing In atoms in various layers. Here we discuss only the most stable structures for each situation (see Table 1); a list of all calculated structures is reported in the Supplemental Material [22].

|  | Fe$_{In}$ at b, In$_{ad}$ at B | Fe$_{ad}$ at B | Fe$_{int}$ |
|---|---|---|---|
| E$_b$ (eV) | 5.00 | 4.37 | 3.67 |
| DE$_b$ (eV) | -0.63 | 0.0 | +0.70 |

Table 1: DFT binding energies of the most stable configurations for Fe incorporation (replacing an In atom), adsorption (ad), and interstitial formation (int). 'b' denotes the central In(6c), atom while 'B' is the adatom adsorption site ontop of it, see Fig. 1.

The most favorable site for the Fe atoms adsorbing on-top of the In$_2$O$_3$(111) surface is site B, i.e., the same site where the In adatoms are most stable on the reduced In$_2$O$_3$(111) surface (see Fig. 1). The interaction of Fe with the In$_2$O$_3$(111) surface is very strong; the binding energy of the Fe atoms amounts to 4.37 eV, which is comparable to the Fe cohesion energy in bulk Fe (experimental formation enthalpy of Fe: 4.31 eV [32]). This may be the reason, why the Fe atoms remain isolated instead of forming clusters. The large binding energy implies that the Fe atoms have a high kinetic energy [33] when they approach the surface and thus, might penetrate into the $In_2O_3(111)$ surface.

All of the probed configurations where Fe atoms occupy interstitial sites below the surface layer are very unfavorable. From our calculations, the most stable interstitial position is more than 0.7 eV less favorable in energy than adsorption on site B i.e., on-top of the surface (Table 1). In two positions it even turned out that the interstitial site is not stable. The Fe atom displaced an In atom from its lattice site, which moved onto the surface and assumed an adatom configuration; this happened spontaneously in the geometry optimization. Finally, all possibilities were probed where an Fe atom replaces an In atom in the surface and subsurface layer, with the respective In atom added on-top of the surface. All configurations are lower in energy than the adsorption of Fe on-top of the surface. In the most favorable configuration the Fe atom sits in the top surface layer at site 'b' directly below the In adatom, which occupies its preferred surface site B, see Fig. 1 (right). The Fe binding energy for this configuration is 5.00 eV, i.e., it is 0.63 eV more favorable than the structure with Fe adatoms in site B, see Fig. 1 (left). However, it does not matter which specific site the Fe substitutes in the topmost layer; in each case, B is the preferred position for the In adatom with one exception: when the Fe atom replaces one of the three other In(6c) atoms surrounding site 'b' (the blue atoms labeled 'd' in Fig. 1). In this case the In adatom moves from site B to an "on-top" position above the Fe; this could represent the source of some disorder in the adatoms as observed in the STM experiments.

To connect these theoretical results with the experiment, we calculated the STM images and the surface corrugation for three situations (see Figs. 3(a-c)): the reduced $In_2O_3(111)$ surface with native In adatoms, the oxidized surface with Fe adatoms, and the most favorable structure, where Fe substitutes for an In atom in the surface layer at site 'b' and the In occupies site B on the surface. These calculations were done with the Bardeen approach using our own post-processing code [30]. Here, the tunneling current is calculated directly from the substrate and the tip wave functions without using any adjustable parameter, in contrast to the commonly used Tersoff-Hamann approximation [34].

As shown in Figs. 3(d-g), the Bardeen calculations predict a much lower corrugation for Fe adatoms than for In adatoms. In contrast, the two configurations with and without substitutional Fe in the surface layer are undistinguishable – independent of the precise position of the Fe atom (see Supplemental Material [22]). The only exception is the case where the In adatom is pulled out of its preferred site B to an "on-top" position above a substituted Fe at site 'd'. Here, the corrugation is still about the same, but the adatom is slightly shifted to a corner of the triangle formed by the In(6c) (see Supplemental Material [22]).

Together with the experimental results, our DFT calculations provide strong support that vapor-deposited Fe atoms penetrate into the $In_2O_3$(111) surface and replace In atoms. In a solid-state redox reaction, atomic Fe is oxidized to $Fe^{3+}$ and reduces the $In_2O_3$ substrate. This is plausible because $Fe_2O_3$ is the more stable oxide in the sense that the reaction

$$2\ Fe_{atom} + (In_2^{3+}O_3^{2-})_{bulk} \rightarrow (Fe_2^{3+}O_3^{2-})_{bulk} + 2\ In_{atom} \qquad (1)$$

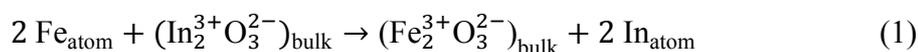

is strongly exothermic with an energy gain of 2.52 eV (calculated from experimental enthalpies [32, 35]).

The most interesting, and a somewhat unexpected aspect, however, is how the system reacts to the reactive incorporation of the excess Fe atoms: The expelled In atoms arrange in a well-ordered pattern on the surface, adopting the same adsorption site as the native In adatoms of the reduced surface. The oxidation state as is nominally assumed by In in $In_2O_3$ is +3 and both, DFT and our XPS measurements (Fig. 2) suggest the same for the incorporated $Fe_{In}$ atoms. In the DFT calculations magnetic moments between 4.7 and 4.9 $\mu_B$ were found for the Fe atoms on In sites, which are typical values for Fe in a +3 oxidation state. However, overall, the material is reduced due to the adatom formation.

Our DFT calculations show that In adatoms are clearly the lowest-energy configuration, much lower than O vacancies, for example. Upon reduction, easily reducible oxides form cation-related defect states (e.g. $Ti^{3+}$ sites in $TiO_2$), in contrast to non-easily reducible oxides, which form O-vacancy-related F-centers (e.g. MgO) [36]. $In_2O_3$ is an easily reducible oxide. The reduction of $In^{3+}$ ions gives rise to a large increase in atomic size. The ionic radius of $In^{3+}$ is 82 pm, whereas for reduced $In^0$ the atomic radius is 162 pm. The change in size is a main driving force why the reduced In ions are expelled from the lattice and form In adatoms. The reduction of $In^{3+}$ ions by occupation of the 5s states is reflected in the electronic structure. In resonant photoemission an In-related gap state was observed upon reduction of Sn-doped $In_2O_3$, and it was shown that this gap state is of In 5s character [37]. A similar gap state appears when we reduce our $In_2O_3$ single crystals [20] and when we deposit Fe (see Supplemental Material [22]). This is in accordance with observations made on another non-transition metal oxide ($SnO_2$, see [38]), where only surface configurations that lead to a $Sn^{2+}$ oxidation state (i.e., a lone pair $5s^2$ configurations resembling SnO) were found to be stable. The similarity of the gap state of these two systems has already been pointed out by Egdell [39].

The regime of ordered adatoms is limited to a coverage that makes a full monolayer, i.e., one Fe atom per surface unit cell. This results in a rather low – and, because of reduced diffusivity of an embedded Fe atom, dilute – distribution of Fe across the near-surface region. (Experimentally, disorder and misplaced adatoms start at even lower coverages due to limited diffusion of the impinging Fe – as well as the expelled In atoms on the surface.) Our data do not permit us to judge at exactly what point the incorporation/push-out stops. XPS (Fig. 2) and STM results (see Supplemental Material [22]) for higher Fe coverages, deposited at RT, are clear indicators that the additional Fe is eventually present in the conventional form, i.e., as excess aggregates in the form of clusters, initially and presumably of a mixed Fe-In composition, and for higher coverages, as Fe nanoclusters.

This reaction is driven by a low cohesive energy of the metal A of the pristine oxide AO, $E_{coh}(A) < E_{coh}(B)$, combined with a more negative formation enthalpy of the new oxide BO, $D_fH^0_{BO} < D_fH^0_{AO}$. Rough estimates based on published thermodynamic data [32, 35] lead us to speculate that a similar process should happen for various other elements; therefor we propose that a similar configuration may evolve for the deposition of other reactive metals such as Ti, Ce, Al, and, also possibly for Ni or Co. When Fe was substituting a lattice In atom, a magnetic moment of Fe between 4.7 and 4.9 $\mu_B$ was found, which are typical values for Fe in a +3 oxidation state.

We propose that the reaction of expelling excess cations to the surface upon reduction is a more general process, and one which may take place on other post-transition metal oxides as well (such as $Ga_2O_3$, for example). In Ref. [18] it was argued that the fact that no reduced phases of $In_2O_3$ are stable, and the relatively low formation enthalpy of In results in excess In on the surface; similar arguments could hold for other materials.

In summary, our study uncovers a novel response of a metal oxide surface to the deposition of a metallic element featuring a thermodynamically favorable incorporation into the surface that is accompanied by a reduction of the system, and – in this case – a response with an array of ordered In adatoms.

**Acknowledgements**

M.W. gratefully acknowledges the FWF project T759-N27. P.L. was supported by the Austrian Science Fund (FWF) within SFB F45 "FOXSI". S.S. thanks the Fonds der Chemischen Industrie (FCI) for a Chemiefonds Fellowship. Research at the Oak Ridge National Laboratory for L.A.B. was sponsored by the U.S. Department of Energy, Basic Energy Sciences, Materials Sciences and Engineering Division. U.D. was supported by the European Research Council Advanced Grant "OxideSurfaces".

**References**

[1] N. Nilius, and H.-J. Freund, Acc. Chem. Res. **48**, 1532 (2015).
[2] J. Repp, G. Meyer, F. E. Olsson, and M. Persson, Science **305**, 493 (2004).
[3] Z. Li, H.-Y. T. Chen, K. Schouteden, K. Lauwaet, L. Giordano, M. I. Trioni, E. Janssens, V. Iancu, C. Van Haesendonck, P. Lievens, and G. Pacchioni, Phys. Rev. Lett. **112**, 026102 (2014).
[4] R. Bliem, J. Pavelec, O. Gamba, E. McDermott, Z. Wang, S. Gerhold, M. Wagner, J. Osiecki, K. Schulte, M. Schmid, P. Blaha, U. Diebold, and G. S. Parkinson, Phys. Rev. B **92**, 075440 (2015).
[5] R. Bliem, E. McDermott, P. Ferstl, M. Setvin, O. Gamba, J. Pavelec, M. A. Schneider, M. Schmid, U. Diebold, P. Blaha, L. Hammer, and G. S. Parkinson, Science **346**, 1215 (2014).
[6] Ch. T. Champbell, Surf. Sci. Rep. **27**, 1 (1997).
[7] U. Diebold, J.-M. Pan, and Th. E. Madey, Surf. Sci. **331**, 845 (1995).
[8] A. Picone, M. Riva, A. Brambilla, A. Calloni, G. Bussetti, M. Finazzi, F. Ciccacci, and L. Duò, Surf. Sci. Rep. **71**, 32 (2016).
[9] H. Onishi, and Y. Iwasawa, Surf. Sci. **313**, L783 (1994).
[10] J. A. Kilner, R. J. Brook, Solid State Ionics **6**, 237-252 (1982).
[11] C. Wagner, Naturwissenschaften **31**, 265 (1943).
[12] S. Halpegamage, P. Ding, X.-Q. Gong, and M. Batzill, ACS Nano **9**, 8627 (2015).
[13] G. Neri, A. Bonavita, G. Micali, G. Rizzo, E. Callone, and G. Carturan, Sensors and Actuators B: Chemical **132**, 224 (2008).
[14] C. Xiangfeng, W. Caihong, J. Dongli, and Z. Chenmou, Chem. Phys. Lett. **399**, 461 (2004).
[15] P. D. C. King T. D. Veal, F. Fuchs, C. Y. Wang, D. J. Payne, A. Bourlange, H. Zhang, G. R. Bell, V. Cimalla, O. Ambacher, R. G. Egdell, F. Bechstedt, and C. F. McConville, Phys. Rev. B **79**, 205211 (2009).
[16] P. D. C. King, T. D. Veal, D. J. Payne, A. Bourlange, R. G. Egdell, and C. F. McConville, Phys. Rev. Lett. **101**, 116808 (2008).
[17] E. H. Morales, Y. He, M. Vinnichenko, B. Delley, and U. Diebold, New J. Phys. **10**, 125030 (2008).

[18] M. Wagner, S. Seiler, B. Meyer, L. A. Boatner, M. Schmid, and U. Diebold, Adv. Mater. Interfaces **1**, 1400289 (2014).
[19] M. Antlanger, W. Mayr-Schmölzer, J. Pavelec, F. Mittendorfer, J. Redinger, P. Varga, U. Diebold, and M. Schmid, Phys. Rev. B **86**, 035451 (2012).
[20] D. R. Hagleitner, M. Menhart, P. Jacobson, S. Blomberg, K. Schulte, E. Lundgren, M. Kubicek, J. Fleig, F. Kubel, C. Puls, A. Limbeck, H. Hutter, L. A. Boatner, M. Schmid, and U. Diebold, Phys. Rev. B **85**, 115441 (2012).
[21] The DFT calculations predict that the energy is lowered by about 8 meV, if the In adatom on the reduced $In_2O_3$(111) surface is slightly displaced from the high-symmetry position B toward one of the three bridging positions between two of the neighboring O(3c) atoms, see Ref. [18]. Since this energy difference is so small, the adatom is expected to move frequently between the three symmetry-equivalent O(3c) bridging positions even at low temperature and can be assumed to be at site B on a time-weighted average.
[22] See Supplemental Material [url], which includes Refs. [23-30].
[23] U. Diebold, H. Tao, N. Shinn, T. Madey, Phys. Rev. B **50**, 14474 (1994).
[24] P. Giannozzi, S. Baroni, N. Bonini, M. Calandra, R. Car, C. Cavazzoni, D. Ceresoli, G.L. Chiarotti, M. Cococcioni, I. Dabo, A. Dal Corso, S. de Gironcoli, S. Fabris, G. Fratesi, R. Gebauer, U. Gerstmann, C. Gougoussis, A. Kokalj, M. Lazzeri, L. Martin-Samos, N. Marzari, F. Mauri, R. Mazzarello, S. Paolini, A. Pasquarello, L. Paulatto, C. Sbraccia, S. Scandolo, G. Sclauzero, A.P. Seitsonen, A. Smogunov, P. Umari, R.M. Wentzcovitch, J. Phys.: Condens. Matter **21**, 395502 (2009).
[25] J.P. Perdew, K. Burke, M. Ernzerhof, Phys. Rev. Lett. 77, 3865 (1996); Erratum: Phys. Rev. Lett. **78**, 1396 (1997).
[26] D. Vanderbilt, Phys. Rev. B **41**, 7892 (1990).
[27] M. Marezio, Acta Crystallographica **20**, 723 (1966).
[28] J. Bardeen, Phys. Rev. Lett. **6**, 57 (1961).
[29] W.A. Hofer, A.S. Foster, A.L. Shluger, Rev. Mod. Phys. **75**, 1287 (2003).
[30] R. Kovacik, B. Meyer, D. Marx, Angew. Chem. Int. Ed. **46**, 4894 (2007).
[31] W. Weiss, and W. Ranke, Prog. in Surf. Sci. **70,** 1 (2002).
[32] M. W. Chase, NIST-JANAF Thermochemical Tables (American Chemical Society, American Institute of Physics for the National Institute of Standards and Technology, Washington, D.C., Woodbury, N.Y., 1998), http://www.worldcat.org/title/nist-janaf-thermochemical-tables/oclc/39682152
[33] U. Diebold, W. Hebenstreit, G. Leonardelli, M. Schmid, and P. Varga, Phys. Rev. Lett. **81**, 405 (1998).
[34] J. Tersoff, and D. R. Hamann, Phys. Rev. B **31**, 805 (1985).
[35] CRC Handbook of Chemistry and Physics, edited by D. R. Lide (CRC Press, Boca Raton, 2010).
[36] L. Giordano, G. Pacchioni, Acc. Chem. Res. **44**, 1244 (2011).
[37] K. H. L. Zhang, R. G. Egdell, F. Offi, S. Iacobucci, L. Petaccia, S. Gorovikov, P. D. King, Phys. Rev. Lett. **110**, 056803-056805 (2013).
[38] M. Batzill, U. Diebold, Prog. Surf. Sci. **79**, 47-154 (2005).
[39] R. G. Egdell, Dopant and defect induced electronic states at $In_2O_3$ surfaces. J. Jupille, G. Thornton (Eds.), Defects at Oxide Surfaces 58, Springer International Publishing (2015) pp. 351–400, DOI 10.1007/978-3-319-14367-5_12.